\documentclass[10pt,a4paper]{article}
\usepackage{amsmath,amssymb,graphicx,geometry,indentfirst,natbib,authblk,bm,aas_macros}
\usepackage[font=small]{caption}
\usepackage[title]{appendix}
\setcitestyle{notesep={}}
\begin{document}
\title{\textbf{Small-Scale Anisotropies of Cosmic Rays from Turbulent Flow}}
\author[1]{Yiran Zhang\footnote{zhangyr@pmo.ac.cn}}
\author[2]{Siming Liu}
\affil[1]{Key Laboratory of Planetary Sciences, Purple Mountain Observatory, Chinese Academy of Sciences, Nanjing 210023, China}
\affil[2]{School of Physical Science and Technology, Southwest Jiaotong University, Chengdu 610031, China}
\maketitle
\begin{abstract}
Within the classical convection--diffusion approximation, we show that the angular distribution of cosmic rays (CRs) in a highly turbulent flow may exhibit significant small-scale anisotropies. The CR intensity angular power spectrum $ C_\ell $ is then a direct reflection of interstellar turbulence, from which one expects $ C_\ell\propto\ell^{-\gamma -1} $ for $ \ell\gg 1 $, where $ \gamma $ is the power-law turbulence spectral index. Observations by IceCube and HAWC at TeV energies can be explained approximately with the Kolmogorov law $ \gamma =5/3 $ with a convection velocity dispersion of 20 km/s on the scale of 10 pc.

\end{abstract}
\section{Introduction}
Cosmic rays (CRs) are mainly relativistic nuclei, with an energy spectral index ($ -\partial\ln f/\partial\ln p-2 $) about 2.7 and relative intensity anisotropy on the order of 0.1\% at TeV energies. The high level of isotropy implies that CRs are diffusive due to uniform pitch-angle scattering by magnetic field irregularities. This is consistent with the fact that the anisotropy is dominated by a dipole signal, which is believed to be associated with the spatial diffusion flux, and can thus be used to trace CR sources \citep{2016PhRvL.117o1103A,2022ApJ...926...41Z,2022MNRAS.511.6218Z,2023ApJ...942...13Q}.

It is not surprising that the observed CR anisotropy deviates from a pure dipole ($ \ell =1 $), but the multipoles with spherical harmonic degrees $ \ell >1 $ cannot be explained by the standard diffusion theory. Besides some exotic scenarios \citep{2013PhLB..725..196K}, most existing models for this issue focus on improving the statistical description of charged particles interacting with a classical electromagnetic field. It has been suggested, e.g., that the basic dipole configuration can be distorted by nonuniform pitch-angle scattering \citep{2010ApJ...721..750M} and/or non-diffusive transport \citep{2016ApJ...822..102H}, which may result from the local magnetic field structure \citep{2012PhRvL.109g1101G,2014Sci...343..988S}.

Interestingly, CR observations by the IceCube experiment and High-Altitude Water Cherenkov (HAWC) Observatory show that the TeV intensity sky map complies with a power-law angular power spectrum $ C_{\ell}\sim C_1\ell ^{-3} $ \citep{2016ApJ...826..220A,2014ApJ...796..108A,2018ApJ...865...57A,2019ApJ...871...96A}. This unique feature could be the key to understand the small-scale anisotropies, and has been interpreted as a consequence of the angular correlation of (two) particle trajectories with similar initial conditions, i.e., the particles undergo so-called relative diffusion in a turbulent magnetic field \citep{2014PhRvL.112b1101A,2015ApJ...815L...2A,2022ApJ...927..110K}.

In this paper, we shall propose an alternative point of view that ascribes $ C_\ell $ to turbulent convection under the classical diffusion approximation. Inspired by \cite{1988ApJ...331L..91E}, in previous work we have emphasized that nonuniform convection can induce dipole and quadrupole ($ \ell =2 $) anisotropies proportional to the diffusion coefficient, via inertial and shear forces, respectively. In particular, the shear anisotropy due to Galactic differential rotation may be important for PeV--EeV CRs \citep{2022ApJ...938..106Z}. This paper aims to extend the convection scenario into the $ \ell >2 $ region.
\section{Particle Back-Tracking}
The basic assumption of the classical convection--diffusion approximation is that particles tend to completely lose information about their initial states due to scattering in the rest frame of the scattering center. In general, the scattering centers at different locations in a flow field have complicated relative motion, i.e., the convection velocity $ \boldsymbol{u} $ is a function of the position vector $ \boldsymbol{r} $. For nonrelativistic convection, the particle momentum in the fluid rest frame is $ \boldsymbol{p}=\boldsymbol{P}-P\boldsymbol{u}\left( \boldsymbol{r} \right) /V $, where $ \boldsymbol{P} $ and $ \boldsymbol{V} $ are the particle momentum and velocity in the observer's frame, respectively. Treating $ \boldsymbol{r} $ and $ \boldsymbol{P} $ as independent canonical variables, when a particle moves between neighboring locations separated by the mean free path $ \boldsymbol{\lambda} $, it obtains a momentum
\begin{align}
\Delta \boldsymbol{p}=\boldsymbol{p}\left( \boldsymbol{r}+\boldsymbol{\lambda} \right) -\boldsymbol{p}\left( \boldsymbol{r} \right) =-\frac{p\Delta \boldsymbol{u}}{v},\label{MI}
\end{align}
where $ \Delta $ denotes the difference between the neighboring systems, and $ \boldsymbol{v} $ is the particle velocity in the fluid rest frame. Note that this expression excludes terms on $ O\left( u^2\right) $ concerning inertial and relativistic corrections for convection.

For brevity, let us adopt the back-tracking argument \citep{2017PrPNP..94..184A,2022ApJ...938..106Z}, which embodies the physics of most interest. We need to work in the fluid rest frame, only in which can complete relaxation due to scattering be observed. Since Liouville's theorem requires statistical fluctuations to be balanced by kinematic terms, if the particle phase-space distribution approaches an isotropic state $ f $ after relaxation (so $ p\partial f/\partial \boldsymbol{p}=\boldsymbol{p}\partial f/\partial p $), the unrelaxed fluctuation (assumed to be small) can be written as the time reversal increment $ \Delta f=\Delta _\text{ICG}f-\boldsymbol{\lambda}\cdot \boldsymbol{\nabla }f $, where the second term corresponds to the diffuse dipole anisotropy, and
\begin{align}
\Delta _\text{ICG}f=-\frac{\partial f}{\partial \boldsymbol{p}}\cdot \Delta \boldsymbol{p}=\frac{\boldsymbol{p}\cdot \Delta \boldsymbol{u}}{v}\frac{\partial f}{\partial p},\label{aniso}
\end{align}
the part associated with nonuniform convection, may be referred to as an inhomogeneous Compton--Getting (CG) effect. The conventional CG effect arises from uniform convection, which yields an additional dipole term $ \Delta _\text{CG}f=\left( -\boldsymbol{P}\cdot \boldsymbol{u}/V \right) \partial f/\partial P $ in the observer's frame. It is worth mentioning that the fluctuation continuously generated by a system under the fluctuation-relaxation equilibrium should be equivalent to $ \Delta f $.

As $ \Delta\boldsymbol{u} $ includes isotropic parts of the fluid deformation (e.g., the isotropic compression or expansion), we have $ \left<\Delta _\text{ICG}f\right> \ne 0 $, where the angle brackets denote averaging over all directions. The fully random fluctuation arising from nonuniform convection should be $ \Delta _\text{ICG}f-\left< \Delta _\text{ICG}f \right> $. See Appx.~\ref{appxa} for a more careful derivation with Bhatnagar--Gross--Krook (BGK) analysis.
\section{The Role of Turbulence}
Eq.~\eqref{aniso} establishes a direct link between the microscopic anisotropy and macroscopic deformation. The small-scale fluctuations are terms associated with the tensor product of multiple directional vectors in the momentum space ($ \boldsymbol{\lambda}/\lambda =\boldsymbol{p}/p $). Therefore, it is nontrivial to expand the system into a Taylor series around $ \boldsymbol{\lambda}=\boldsymbol{0} $,
\begin{align}
\Delta \boldsymbol{u}=\sum_{l=2}^{\infty}{\boldsymbol{U}_{l-1}}=\sum_{l=2}^{\infty}{\frac{\left( \boldsymbol{\lambda}\cdot \boldsymbol{\nabla } \right) ^{l-1}\boldsymbol{u}}{\left( l-1 \right) !}},
\end{align}
where $ \boldsymbol{U}_{l-1} $ determines the $ l $th term in the corresponding expansion of $ \Delta _\text{ICG}f $. On the other hand, $ \Delta _\text{ICG}f $ can also be written as a spherical harmonic expansion, whose $ \ell $th term defines the irreducible $ 2^\ell $-pole anisotropy. The two expansions are generally inequivalent (see Appx.~\ref{appxa}; note the distinction between $ l $ and $ \ell $), while CR anisotropies are expressed via the spherical harmonics.

Here, the significance of the Taylor expansion is that it facilitates our understanding to the role of fluid regularity. In principle, the fluid description requires $ \Lambda\gtrsim\lambda $ such that the local system of particles can be relaxed, where $ \Lambda $ is the macroscopic scale on which the flow property changes significantly. Assuming $ u\propto\Lambda ^\nu $, we make the following order-of-magnitude estimate,
\begin{align}
U_{l-1}\sim \frac{1}{\left( l-1 \right) !}\left( \lambda\frac{\partial}{\partial \Lambda} \right) ^{l-1}u=\frac{u\Gamma \left( \nu +1 \right)}{\Gamma \left( \nu +2-l \right) \Gamma \left( l \right)}\left( \frac{\lambda}{\Lambda} \right) ^{l-1},
\end{align}
where $ \Gamma $ denotes the gamma function. This reduces to $ U_l\sim u\Gamma \left( \nu +1 \right) \left( \lambda /\Lambda \right) ^l\sin \left[ \pi \left( \nu -l \right) \right] /\left( \pi l^{\nu +1} \right) $ when $ l\gg 1 $.

The factor $ \left( \lambda /\Lambda \right) ^l $ implies that the large-$ l $ corrections of $ \Delta _\text{ICG}f $ are not important for $ \Lambda >\lambda $. Considering that $ \lambda $ typically increases with the particle energy, in a regular nonuniform flow there should be no significant $ l>2 $ anisotropy arising from convection, except for the most energetic particles with $ \lambda\sim\Lambda $. However, CR small-scale anisotropies are observed at different energies. To avoid the high-$ l $ cutoff over a broad spectrum, we need to consider a highly turbulent flow, which varies irregularly between any two spatial points so that $ \Lambda\sim\lambda $ can be maintained at any energy. It can be seen from the detracing projection \citep{1989JPhA...22.4303A} that the $ l $th Taylor term is composed of $ \left\{2^{l-2n}\mid 0\leqslant 2n\leqslant l,n\in\mathbb{Z}\right\} $-pole structures, hence we are led to conclude that turbulence is capable of producing polyenergetic multipole anisotropies that do not exponentially decay with $ \ell $.

One may ask: Since structures with $ \Lambda <\lambda $ are hydrodynamically unresolvable, how can such small-scale anisotropies have a fluid description? To answer this, it is necessary to distinguish between fore- and back-ground flows: the former represents the flow of particles under consideration, while the latter is that of scattering centers. The establishment of relaxation means that the two flows share the same $ \boldsymbol{u} $ on the scale $ \lambda $, whereas the background flow is also defined on smaller scales. It is $ \Delta\boldsymbol{u} $, which can now be interpreted as the background nonuniformity, and thus as a superposition of the background small-scale structures, that determines the configuration of the foreground small-scale anisotropies.

The spectrum of the structures is encoded in the Fourier expansion of the velocity correlation function. But we are more interested in the multipole expansion
\begin{align}
\Delta \boldsymbol{u}=\sum_{L=0}^{\infty}{\boldsymbol{u}_L}=\sum_{L=0}^{\infty}{\sum_{M=-L}^L{\boldsymbol{c}_{LM}Y_{LM}\left( \frac{\boldsymbol{\lambda}}{\lambda} \right)}},\label{ME}
\end{align}
where $ Y_{LM} $ is the spherical harmonic function. Using the plane-wave expansion \citep{1991qmnr.book.....L}, one has $ \boldsymbol{c}_{LM}=4\pi i^L\int{j_L\left( k\lambda \right) Y_{LM}^*\left( \boldsymbol{k}/k \right) \tilde{\boldsymbol{u}}\left( \boldsymbol{k} \right) d^3k/\left( 2\pi \right) ^3} $, where $ j_L $ refers to the spherical Bessel function of the first kind, and $ \tilde{\boldsymbol{u}} $ is the Fourier transform of $ \Delta\boldsymbol{u} $.

Let the overline represent an ensemble average over all possible realizations of the local field. As the two-point correlation $ \overline{\Delta \boldsymbol{u}\left( \boldsymbol{\lambda }_1 \right) \Delta \boldsymbol{u}\left( \boldsymbol{\lambda }_2 \right) } $ is a function of $ \boldsymbol{\lambda}_1-\boldsymbol{\lambda}_2 $, $ \overline{\tilde{\boldsymbol{u}}\left( \boldsymbol{k} \right) \tilde{\boldsymbol{u}}^*\left( \boldsymbol{k}' \right) } $ must contain the Dirac delta function $ \delta ^3\left( \boldsymbol{k}-\boldsymbol{k}' \right) $ \citep{1987flme.book.....L}. For convenience, let us consider an equipartition system with $ \overline{\tilde{u}_{\alpha}\left( \boldsymbol{k} \right) \tilde{u}_{\alpha '}^{*}\left( \boldsymbol{k}' \right) }=\left( 2\pi \right) ^6\delta ^3\left( \boldsymbol{k}-\boldsymbol{k}' \right) \delta _{\alpha \alpha '}w/\left( 12\pi k^2 \right) $, where $ \delta_{\alpha \alpha '} $ is the Kronecker delta ($ \alpha $ and $ \alpha '=x $, $ y $ or $ z $), and $ w $ is the omnidirectional turbulence spectrum. Then, from the orthogonality condition $ 4\pi \left< Y_{LM}Y_{L'M'}^* \right> =\delta _{LL'}\delta _{MM'} $ and Uns\"old’s theorem \citep{2022ApJ...927..110K}, we have
\begin{align}
\overline{u_{L}^{2}}=\overline{\left< u_{L}^{2}\right>}=\left( 2L+1 \right) \int_0^{\infty}{j_{L}^{2}\left( k\lambda \right) w\left( k \right) dk},\label{TS}
\end{align}
and $ \overline{c_{LM\alpha}c_{L'M'\alpha '}^{*}}=4\pi\overline{u_{L}^{2}}\delta _{LL'}\delta _{MM'}\delta _{\alpha \alpha '}/\left[ 3\left( 2L+1 \right) \right] $, implying that the turbulence energy is shared equally among all directional degrees of freedom. Note the identity $ \sum_{L=0}^{\infty}{\left( 2L+1 \right) j_{L}^{2}}=1 $.

Because $ j_L\left( \xi\right) $ peaks around $ \xi =L+\chi_L $ with $ \lim_{L\rightarrow \infty} \chi_L/L=0 $ \citep[see, e.g.,][, who used $ \chi _L=1/2 $]{2023PhRvD.108j3007F}, and $ \int_0^{\infty}{j_{L}^{2}\left( \xi \right) d\xi}=\pi /\left[ 2\left( 2L+1 \right) \right] $, we may take the approximation $ j_{L}^{2}\left( \xi \right) \sim \pi \delta \left( \xi -L-\chi _L \right) /\left[ 2\left( 2L+1 \right) \right] $, leading to $ \overline{u_{L}^{2}}\sim \pi w\left( k_L \right) /\left( 2\lambda \right) $ with $ k_L\lambda =L+\chi _L $. In this regard, there is a simple geometric interpretation: Structures described by the $ L $th term of Eq.~\eqref{ME} have an effective angular scale $ 2\pi /\left( L+\chi _L\right) $, corresponding to two points separated by a length scale $ 2\pi\lambda /\left( L+\chi _L\right) $ on a sphere of radius $ \lambda $, while in the interval $ k_{L+1}-k_L $ the velocity variance between the two points is roughly $ \overline{u_{L}^{2}} $. So the dependence of $ \overline{u_{L}^{2}} $ on $ L $ is similar to that of $ w\left( k_L\right) $ on $ k_L $; whereas they are different by a smoothing factor since the width of $ j_{L}^{2} $ is actually nonzero.

As an example, the exact result of Eq.~\eqref{TS} for a power-law spectrum $ w\propto k^{-\gamma} $ with $ \gamma <2L+1 $ is
\begin{align}
\overline{u_{L}^{2}}=\frac{\sqrt{\pi}\left( 2L+1 \right) w\left( \frac{1}{\lambda} \right) \Gamma \left( \frac{\gamma +1}{2} \right) \Gamma \left( L+\frac{1-\gamma}{2} \right)}{4\lambda \Gamma \left( \frac{\gamma}{2}+1 \right) \Gamma \left( L+\frac{\gamma +3}{2} \right)}\label{PL}
\end{align}
($ \gamma =\text{const} $). This reduces to $ \overline{u_{L}^{2}}=\sqrt{\pi}w\left( k_L \right) \Gamma \left[ \left( \gamma +1 \right) /2 \right] /\left[ 2\lambda \Gamma \left( \gamma /2+1 \right) \right] $ in the large-$ L $ limit.
\section{Multipole Anisotropies}
According to Eqs.~\eqref{aniso} and \eqref{ME}, the $ 2^\ell $-pole moment of $ \Delta _\text{ICG}f $ is determined by the dipole--$ 2^L $-pole coupling
\begin{align}
\frac{\boldsymbol{\lambda}\cdot \Delta \boldsymbol{u}}{\lambda}=\sum_{\ell =0}^{\infty}{\sum_{m=-\ell}^{\ell}{a_{\ell m}Y_{\ell m}^{*}}}=\sum_{L=0}^{\infty}{\sum_{M=-L}^L{Y_{LM}\sum_{N=-1}^1{b_{LMN}Y_{1N}}}},
\end{align}
where $ b_{LM,\pm 1}=\sqrt{2\pi /3}\left( ic_{LMy}\mp c_{LMx} \right) $, $ b_{LM0}=\sqrt{4\pi /3}c_{LMz} $, and
\begin{align}
a_{\ell m}=\sqrt{\frac{3\left( 2\ell +1 \right)}{4\pi}}\sum_{L=0}^{\infty}{\sqrt{2L+1}\left( \begin{matrix}
	L&		1&		\ell\\
	0&		0&		0\\
	\end{matrix} \right) \sum_{M=-L}^L{\sum_{N=-1}^1{b_{LMN}\left( \begin{matrix}
			L&		1&		\ell\\
			M&		N&		m\\
			\end{matrix} \right)}}}.
\end{align}
Selection rules of the Wigner 3j symbols \citep{1991qmnr.book.....L} require the nonzero terms to satisfy $ L=\left| \ell\pm 1 \right| $ and $ M=-m-N $. Since $ L=0 $ and $ \ell =0 $ both represent mean fields, in general only anisotropies with $ \ell\geqslant 2 $ can be ascribed to pure turbulent convection.

The steady-state angular correlation of the anisotropies exists due to nonvanishing $ \overline{a_{\ell m}a_{\ell 'm'}^*} $, which is controlled by $ \overline{\boldsymbol{c}_{LM}\boldsymbol{c}_{L'M'}^*} $. On adopting the equipartition hypothesis (see Eq.~\ref{TS}), we have $ \overline{b_{LMN}b_{L'M'N'}^{*}}=\left( 4\pi /3 \right) ^2\overline{u_{L}^{2}}\delta _{LL'}\delta _{MM'}\delta _{NN'}/\left( 2L+1 \right) $. It follows that $ \overline{a_{\ell m}a_{\ell 'm'}^{*}}=\left( A_{\ell}\delta _{\ell \ell '}+\mathcal{A}_{\ell m}\delta _{\ell ,\ell '+2}+\mathcal{A}_{\ell 'm}\delta _{\ell ',\ell +2} \right) \delta _{mm'} $, while the angular power is determined by the diagonal elements
\begin{align}
A_{\ell}=\frac{4\pi}{3\left( 2\ell +1 \right)}\left( \frac{\ell}{2\ell -1}\overline{u_{\ell -1}^{2}}+\frac{\ell +1}{2\ell +3}\overline{u_{\ell +1}^{2}} \right) ,\label{Al}
\end{align}
which are $ m $-independent. Therefore, any dependence of $ \overline{\left| a_{\ell m} \right|^2} $ on $ m $ would imply that the equipartition theorem is broken in an asymmetric system.

By convention \citep{2017PrPNP..94..184A}, we define $ 4\pi \overline{\left< \left( \Delta _{\text{ICG}}f \right) ^2 \right> }/f^2=\sum_{\ell =0}^{\infty}{\left( 2\ell +1 \right) C_{\ell}} $, where
\begin{align}
C_{\ell}=\left( \frac{1}{v}\frac{\partial \ln f}{\partial \ln p} \right) ^2\frac{1}{2\ell +1}\sum_{m=-\ell}^{\ell}{\overline{\left| a_{\ell m} \right|^2}}\label{PS}
\end{align}
is the steady-state angular power spectrum of the relative anisotropy induced by nonuniform convection. Note that a power-law spectrum of turbulence (see Eq.~\ref{PL}) yields $ C_\ell\propto\ell ^{-\gamma -1} $ for $ \ell\gg 1 $.
\section{Data Fit}
\begin{figure}
	\centering
	\includegraphics[width=1\textwidth]{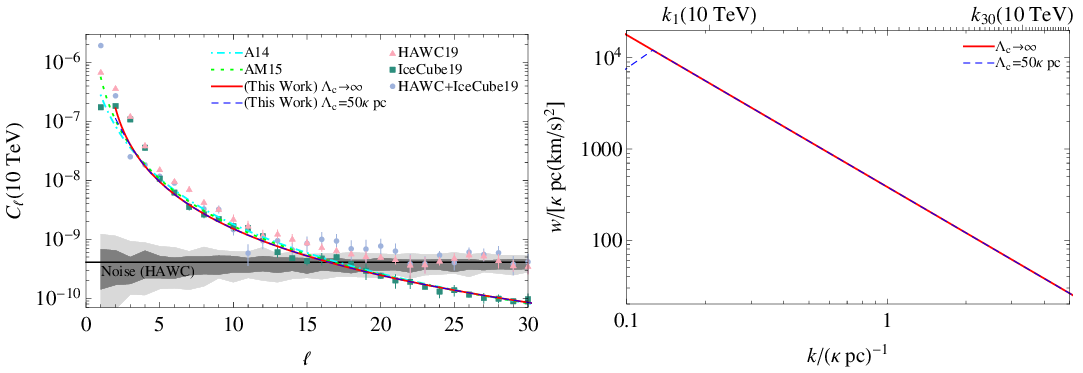}
	\caption{Left: Fits to the CR angular power spectrum at 10 TeV, with the observational data from \cite{2019ApJ...871...96A}. The red solid and blue long-dashed curves represent scenarios described by Eqs.~\eqref{TS}, \eqref{Al} and \eqref{PS}. The cyan dot-dashed curve marks the \citet[; A14]{2014PhRvL.112b1101A} model. The green short-dashed curve refers to modeling by \citet[; AM15]{2015ApJ...815L...2A}. The black solid line shows the noise level dominated by HAWC, with the dark and light gray bands indicating the 68\% and 95\% confidence levels, respectively. Though the noise level for IceCube alone is not displayed in \cite{2019ApJ...871...96A}, from the sky coverage and number of events one can estimate it to be around $ 10^{-11} $ \citep{2017PrPNP..94..184A}. Right: Turbulence spectra used in the calculation. Note that here we set $ j_L\left( k_L\lambda \right)=\max j_L $ for labeling $ k_L $.}\label{f1}
\end{figure}
It is mathematical and physical simplicity that motivates us to consider the equipartition hypothesis. Unfortunately, moduli of spherical harmonic coefficients of the TeV CR intensity have been reported to be $ m $-dependent \citep[see Tab.~3 in][]{2019ApJ...871...96A}. Nevertheless, we notice that the coefficient of variance for such moduli at any fixed $ \ell $ is smaller than 1. Moreover, it remains possible that the observing time is insufficient. On the other hand, as Eq.~\eqref{PS} washes out the dependence on $ m $, there must be an effective equipartition flow that produces the same $ C_{\ell} $. For simplicity, let us assume the effective flow to be an acceptable approximation to the realistic one; especially, they have similar direction-averaged spectra.

In Fig.~\ref{f1}, we show that the observed CR angular power spectrum at a median energy of 10 TeV can be described for $ \ell >1 $ with the above turbulent convection scenario, which fulfills the Kolmogorov law $ \gamma =5/3 $. The $ \ell =1 $ term is not included in our calculation because it involves the regular field $ \boldsymbol{u}_0 $ (see Eq.~\ref{Al}). Given the higher noise level of HAWC, we fit only the IceCube data to cover the highest multipole moment provided ($ \ell =30 $), while the two experiments indicate basically the same spectral shape for $ 1<\ell\lesssim 15 $. Assuming $ \partial\ln f/\partial\ln p =-4.7 $, the data fit with Eq.~\eqref{PL} (i.e., $ \Lambda_\text{c}\rightarrow\infty $) gives $ \sigma \left( 10\text{ TeV} \right) \approx 23 $ km/s, where $ \sigma =\sqrt{2\pi w\left( 2\pi /\lambda \right) /\lambda} $ may be recognized as the CR convection velocity dispersion. For comparison, we also display the fitting results from other two standard scenarios concerning relative diffusion \citep{2014PhRvL.112b1101A,2015ApJ...815L...2A}, whose analytic expressions can be found in Appx.~\ref{appxb}.

From Galactic CR propagation models, we learn $ \lambda\left( 10\text{ TV}\right) \sim 10\kappa$ pc, where $ \kappa\sim 1 $ is the diffusion coefficient (defined as $ v\lambda /3 $) at 10 TV in units of $ 3\times 10^{29}\text{ cm}^2 $/s \citep{2022MNRAS.511.6218Z}. Therefore, taking into account the fact that CRs in the TeV range are mainly protons, $ \ell =2 $ (determined mainly by $ L=1 $) and 30 at 10 TeV correspond roughly to scales of $ 20\pi \kappa /\left( 1+\chi _1 \right) $ and $ 20\pi \kappa /\left( 30+\chi _{30} \right) $ pc, respectively ($ 1\sim \chi _1<\chi _{30}\ll 30 $; see the aforementioned delta-function approximation of $ j_L^2 $). Observations have inferred that there is indeed Kolmogorov-like turbulence of the interstellar medium (ISM) on scales around 2--30 pc \citep{2010ApJ...710..853C,2019NatAs...3..154L,2020ApJ...904...66L}, in agreement with our convection scenario. However, the observed ISM velocity dispersion on the scale of 10 pc is typically only several km/s, even for ionized gases \citep{2022ApJ...934....7H}.
\begin{figure}
	\centering
	\includegraphics[width=1\textwidth]{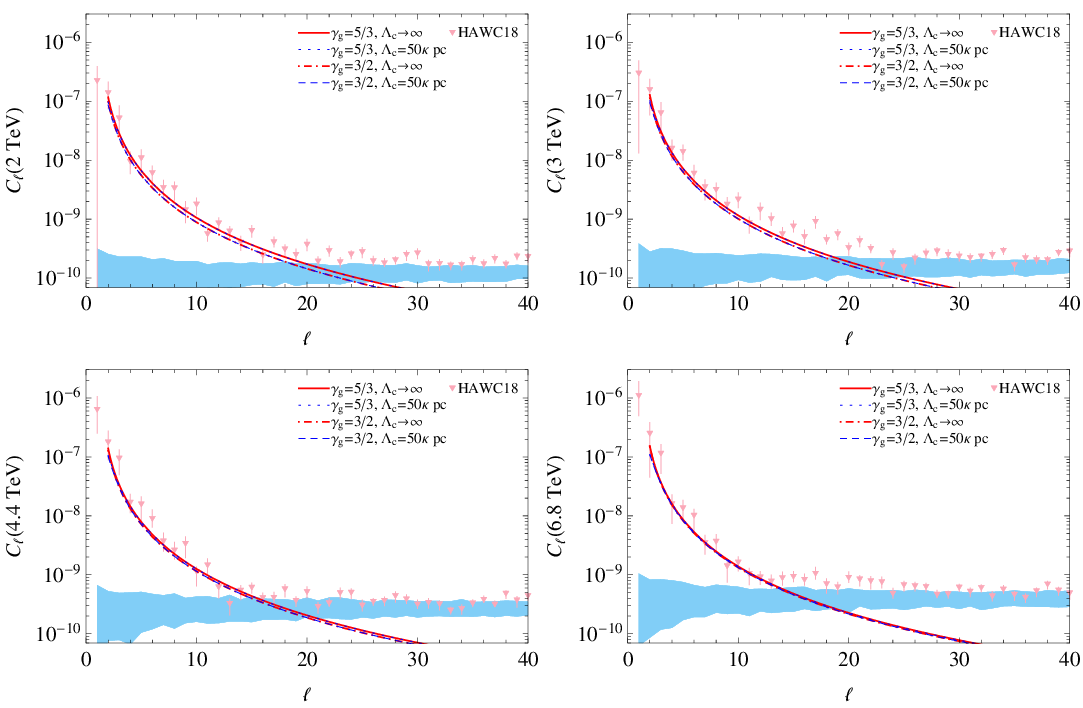}
	\caption{Multi-energy $ C_\ell $ corresponding to fitting results in Fig.~\ref{f1}, with the observational data from \citet[; including systematic errors]{2018ApJ...865...57A}. The colored bands show the noises at a 90\% confidence level.}\label{f2}
\end{figure}

In fact, scattering centers of CRs are more likely to be plasma waves, whose propagation may be characterized with the Alfv\'en velocity $ \boldsymbol{v}_{\text{A}}=\boldsymbol{B}/\sqrt{4\pi\rho} $, where $ \boldsymbol{B} $ is the magnetic field, and $ \rho $ is the mass density. Propagation of uncertainty yields $ \left( \sigma _{\boldsymbol{v}_{\text{A}}}/\left| \overline{\boldsymbol{v}_{\text{A}}} \right| \right) ^2=\left( \sigma _{\boldsymbol{B}}/\left| \overline{\boldsymbol{B}} \right| \right) ^2+\left[ \sigma _{\rho}/\left( 2\overline{\rho } \right) \right] ^2 $, with $ \sigma _{\boldsymbol{v}_{\text{A}}} $, $ \sigma _{\boldsymbol{B}} $ and $ \sigma _{\rho} $ the corresponding root-mean-square values. On the other hand, the quasilinear diffusion theory claims $ \lambda \sim r_{\text{g}}\left[ \left| \overline{\boldsymbol{B}} \right|/\sigma _{\boldsymbol{B}}\left( 2\pi r_{\text{g}} \right) \right] ^2 $, where $ r_\text{g} $ is the gyroradius, and $ \sigma _{\boldsymbol{B}}^2\left( 2\pi r_{\text{g}} \right) =\sigma _{\boldsymbol{B}}^2\left( \lambda \right) \left( 2\pi r_{\text{g}}/\lambda \right) ^{\gamma _\text{g}-1} $ is the mean-square magnetic field at the resonance wavenumber $ 1/r_\text{g} $ with $ \gamma _\text{g} $ the power-law spectral index of turbulence on the same scale \citep{2007ARNPS..57..285S}. Setting $ \lambda =10 $ pc, $ r_\text{g}=0.003 $ pc and $ \gamma _\text{g}=\gamma =5/3 $, for the local ISM with $ \left| \overline{\boldsymbol{B}} \right|\sim 3\ \mu $G and $ \overline{\rho}\sim 0.5\sigma _{\rho}\left( 10\text{ pc}\right)\sim 1.67\times 10^{-25}\text{ g}/\text{cm}^3 $ \citep{2019NatAs...3..154L} we obtain $ \sigma _{\boldsymbol{v}_{\text{A}}}\left( 10\text{ pc}\right)\sim 20 $ km/s, which is comparable with the fitted $ \sigma \left( 10\text{ TeV} \right) $.
\begin{figure}
	\centering
	\includegraphics[width=1\textwidth]{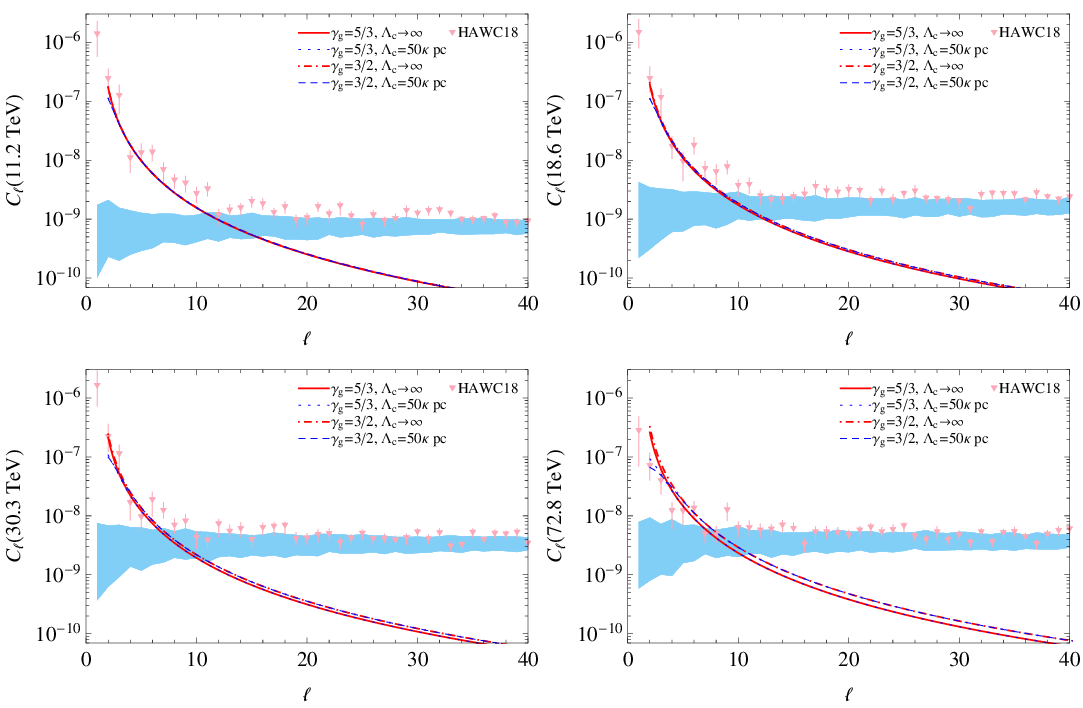}
	\caption{Multi-energy $ C_\ell $ (continued from Fig.~\ref{f2}).}\label{f3}
\end{figure}

Consequently, we expect $ C_\ell\left( 10\text{ TeV}\right)\propto\ell ^{-8/3} $ for $ \ell\gg 1 $ until $ k_L\left( 10\text{ TeV}\right) $ reaches a broken wavenumber, around which the turbulence spectral index changes. Indeed, observational and theoretical uncertainties still admit a moderate variation of the index, while any scale dependence of the index must be imprinted on $ C_\ell $. For example, as it has been claimed that the turbulence power may be enhanced (with respect to a single power-law spectrum) below the heliospheric scale of $ 2\pi\eta\times 120 $ AU \citep[$ \eta\sim 1 $;][]{2019ApJ...887..116Z,2020ApJ...904...66L}, we accordingly predict an enhancement of $ C_\ell $ for $ \ell \gtrsim \lambda /\left( 120\eta\text{ AU} \right) $, which is however far beyond the resolution of existing TeV observations.

Since the diffusion theory indicates $ \lambda\propto p^{2-\gamma _\text{g}} $, meanwhile $ \overline{u_L^2}\propto\lambda ^{\gamma -1} $ as seen from Eq.~\eqref{PL}, Eq.~\eqref{Al} implies $ C_\ell\propto p^{\left( \gamma -1 \right) \left( 2-\gamma _\text{g} \right)} $. We emphasize that this energy dependence is valid for $ k_L $ far away from the broken wavenumbers; and under large-scale Kolmogorov turbulence ($ \gamma =5/3 $), one has $ C_\ell\propto p^{2/9} $ or $ p^{1/3} $ for CRs gyroresonantly scattered by small-scale magnetic irregularities ($ k_Lr_\text{g}\ll 1 $) with a Kolmogorov ($ \gamma _\text{g}=5/3 $) or Kraichnan ($ \gamma _\text{g}=3/2 $) spectrum, respectively.

\begin{figure}[ht]
	\centering
	\includegraphics[width=1\textwidth]{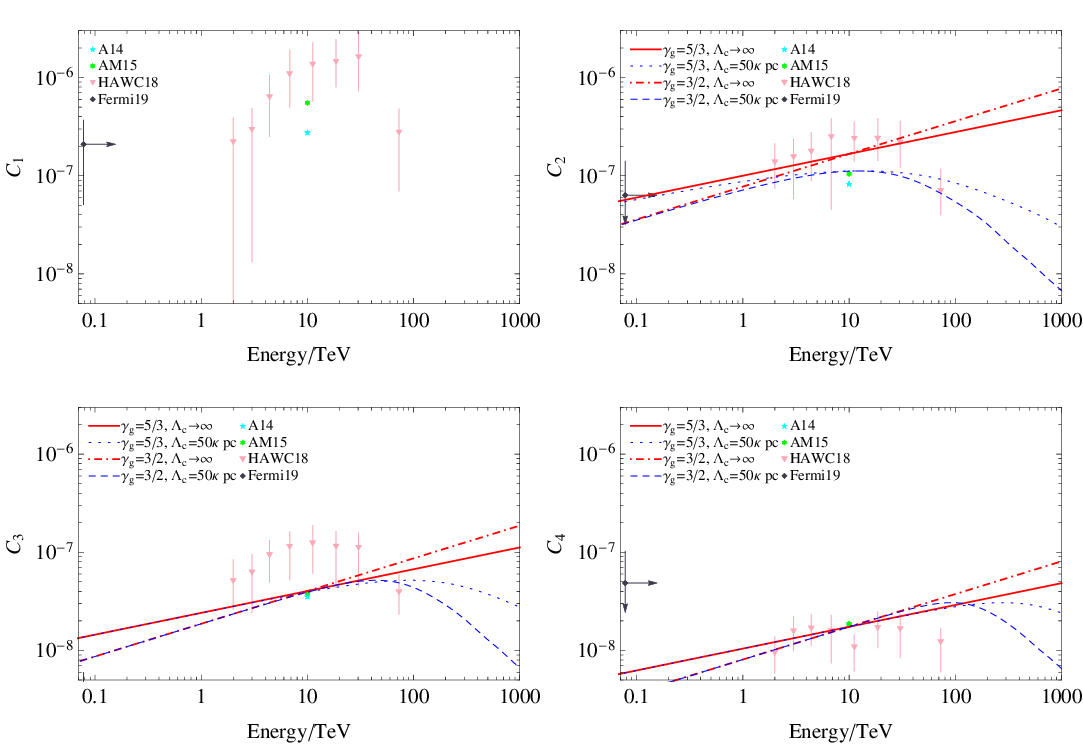}
	\caption{Energy dependence of some $ C_\ell $ (continued from Figs.~\ref{f2} and \ref{f3}), where the Fermi data, which appear to be mainly fluctuations around 0 when $ \ell >2 $, are from \citet[; noise-subtracted]{2019ApJ...883...33A}.}\label{f4}
\end{figure}

With the same $ \gamma $ and $ \sigma\left( 10\text{ TeV}\right) $ mentioned before, we plot energy variations of $ C_\ell $ in Figs.~\ref{f2}--\ref{f4}. As seen, although turbulent convection with a power-law spectrum can provide a rough description to the angular power spectra measured by HAWC (above the noise level) at multi-TeV energies, there is a tendency that the observed energy dependence in 10--100 TeV deviates from the power law below 10 TeV.

Can this ``anomaly'' be ascribed to the scale dependence of the turbulence spectral index? Intuitively, there are two scenarios for the amplitude decrease with energy in 10--100 TeV: (i) $ \gamma >1 $ and $ \gamma _\text{g}>2 $, which may partly account for efficient damping of plasma waves on scales between $ 2\pi r_\text{g}\left( 10\text{ TeV}\right) \sim 0.02 $ pc and $ 2\pi r_\text{g}\left( 100\text{ TeV}\right) \sim 0.2 $ pc; (ii) $ \gamma <1 $ and $ \gamma _\text{g}<2 $, which may partly arise from energy injection by supernova remnants (SNRs) at a correlation length $ \Lambda_\text{c}\sim 50 $ pc.

In the following, we shall quantify only the injection effect from SNRs, and leave other effects for future work. For simplicity, let us consider the steady-state solution of the wave diffusion equation with monochromatic injection \citep{1990JGR....9514881Z}, i.e., $ w\propto \left[ \Lambda _{\text{c}}k/\left( 2\pi \right) \right] ^2\Theta \left( 2\pi /\Lambda _{\text{c}}-k \right) +\left[ \Lambda _{\text{c}}k/\left( 2\pi \right) \right] ^{-\gamma}\Theta \left( k-2\pi /\Lambda _{\text{c}} \right) $, where $ \Theta $ denotes the unit step function. The results in Fig.~\ref{f4} suggest that the anomaly cannot be explained with the SNR injection effect alone, because, most importantly, the data exhibit no strong $ \ell $-dependence of the critical momentum $ p_\text{c} $, around which the energy dependence of $ C_\ell $ changes, while the theory predicts $ p_{\text{c}}\propto \ell ^{1/\left( 2-\gamma _{\text{g}} \right)} $ according to $ k_L\left( p_{\text{c}} \right) \sim 2\pi /\Lambda _{\text{c}} $. Nevertheless, spectral flattening of SNR-driven turbulence on tens-of-pc scales is of physical significance \citep{2010ApJ...710..853C}. In addition, the $ \ell $-independent $ p_\text{c} $ may reflect a change in the energy dependence of $ \lambda $, which is controlled by $ w\left( 1/r_\text{g}\right) $.

Above 100 TeV, existing observations indicate that the energy dependence of the anisotropy amplitude projected onto the equatorial plane becomes a ``standard'' power law again, but the normalization factor is significantly smaller than that below 10 TeV \citep{2022ApJ...938..106Z}. Recently, similar results of the angular power have also been reported preliminarily by IceCube using 11 years of data \citep{2023arXiv230802331M}. We accordingly speculate that interstellar turbulence has a low-level component (compared with the SNR-driven one), which is injected due to Galactic large-scale convection at a correlation length no less than kpc.
\section{Summary}
With the idea that particle distributional fluctuations can be related to fluid local nonuniformity, we have demonstrated that turbulent convection may give rise to remarkable small-scale anisotropies of the distribution. Of particular note is that this scenario is based on the classical convection--diffusion approximation, which is ruled out by other models for the CR small-scale anisotropy problem.

Our model, described by Eqs.~\eqref{TS}, \eqref{Al} and \eqref{PS}, indicates that the $ 2^\ell $-pole anisotropy total power $ \left( 2\ell +1\right) C_\ell $ is roughly proportional to $ w\left( \ell/\lambda\right) /\lambda $, the variance of the turbulent velocity at the scale $ 2\pi\lambda /\ell $ in the characteristic wavenumber interval $ 1/\lambda $. A power-law spectrum $ w\propto k^{-\gamma} $ exactly leads to $ C_\ell\propto\ell^{-\gamma -1} $ for $ \ell\gg 1 $. The fit to observational data brings an attractive picture that interstellar turbulence, with the Kolmogorov law $ \gamma =5/3 $ valid at least on scales of 2--30 pc, and a velocity dispersion about 20 km/s for Alfv\'enic convection on the scale of 10 pc, is responsible for CR multipole anisotropies in the TeV range. This hypothesis remains to be further tested by future precise measurements of broad-spectrum anisotropies and turbulence.
\section*{Acknowledgment}
This work is supported by grants from the National Natural Science Foundation of China (Nos.~U1931204, 12303055, 12375103), National Key R\&D Program of China (2018YFA0404203), and Jiangsu Funding Program for Excellent Postdoctoral Talent (2022ZB476).
\begin{appendices}
\counterwithin{equation}{section}
\renewcommand{\theequation}{\Alph{section}\arabic{equation}}
\section{BGK Approximation}\label{appxa}
For the sake of simplicity, let us consider a steady-state fluctuation--relaxation system of particles under pure scattering (i.e., in the absence of ``external'' forces). In the observer's frame, this can be described with the Boltzmann equation
\begin{align}
\boldsymbol{V}\cdot \boldsymbol{\nabla }\left( f+\Delta f \right) =-\frac{\Delta f}{T},\label{BE}
\end{align}
where the notations are the same as those in the main text, and $ T $ is the relaxation time. To derive the equation in the fluid rest frame, in which the particle distribution tends to be isotropized, it is necessary to introduce a nonuniform momentum transformation, with space-time coordinates fixed. Accordingly, the phase-space gradient should be transformed via the chain rule. Since the Lorentz transformation is allowed at each location, the distribution function is invariant, and time dilation leads to $ T=\lambda P/\left( Vp \right) $. After substituting these transformations, and subtracting $ \left<\text{Eq.~}\eqref{BE}\right> $ from Eq.~\eqref{BE}, one can express the random perturbation $ \Delta f $ in terms of the isotropic distribution $ f $, and then arrive at the standard convection--diffusion equation \citep{1988ApJ...331L..91E,1989ApJ...340.1112W}.

Mathematically, it can be seen that nonuniform convection enters the system via the spatial gradient transformation. The usual chain rule employs only first-order derivatives, including $ \boldsymbol{\nabla u} $ that gives rise to inertial and shear effects. To investigate effects on smaller scales, we take into account the fact that the fluid system has a minimal resolution scale $ \lambda $. Since the flow field is now defined on a lattice, it seems necessary to introduce some finite differences of $ \boldsymbol{u} $. On the other hand, Hamilton's equations formally require the gradient transformation to be carried out between first-order derivatives of the distribution function with respect to kinematic variables. Consequently, it may be reasonable to take
\begin{align}
\boldsymbol{\nabla }_{\boldsymbol{P}}=\boldsymbol{\nabla }_{\boldsymbol{p}}+\left( \boldsymbol{K}\Delta \boldsymbol{p} \right) _{\boldsymbol{P}}\cdot \frac{\partial}{\partial \boldsymbol{p}},\label{DT}
\end{align}
where the subscripts (usually omitted for brevity) emphasize that the operations are under constant $ \boldsymbol{P} $ or $ \boldsymbol{p} $, and $ \boldsymbol{K} $ is the reciprocal free path satisfying $ K_\alpha\lambda_{\alpha '}=\delta _{\alpha\alpha '} $. From the Taylor expansion, we learn $ \boldsymbol{\lambda}\cdot \left( \boldsymbol{K}\Delta \boldsymbol{p} \right) =\Delta \boldsymbol{p} $. Note that the conventional chain rule can be recovered with $ \Delta\boldsymbol{p}=\boldsymbol{\lambda}\cdot \boldsymbol{\nabla p} $.

When averaging over solid angles, it is natural to consider that $ \Delta\boldsymbol{u} $ (see Eq.~\ref{MI}) contains entire information about the asymmetry of the local fluid element, while $ \boldsymbol{u} $ is direction-independent. Neglecting terms on $ O\left( u^2\right) $ and $ O\left( u\Delta f\right) $, applying $ \left<\Delta f \right> =0 $, Eq.~\eqref{DT} and the momentum transformation to $ \left<\text{Eq.~}\eqref{BE}\right> $ yields
\begin{align}
\boldsymbol{u}\cdot \boldsymbol{\nabla }f-\frac{\left< \boldsymbol{p}\cdot \Delta \boldsymbol{u} \right>}{\lambda}\frac{\partial f}{\partial p}=-\boldsymbol{\nabla }\cdot \left< \boldsymbol{v}\Delta f \right> .\label{AE}
\end{align}
Subtracting it from Eq.~\eqref{BE}, with assuming $ \Delta f $ to be spatially nearly homogeneous, or treating $ \left< \boldsymbol{v}\Delta f \right> $ as the leading-order term in the perturbation expansion of $ \boldsymbol{v}\Delta f $, we obtain
\begin{align}
\Delta f\sim \frac{\boldsymbol{p}\cdot \Delta \boldsymbol{u}-\left< \boldsymbol{p}\cdot \Delta \boldsymbol{u} \right>}{v}\frac{\partial f}{\partial p}-\boldsymbol{\lambda}\cdot \boldsymbol{\nabla }f.
\end{align}
This can be substituted back into Eq.~\eqref{AE} to derive an isotropic transport equation, which is beyond the scope of this paper.

The Cartesian description of the convectional small-scale anisotropies consists in the Taylor expansion
\begin{align}
\boldsymbol{\lambda }\cdot \Delta \boldsymbol{u}-\left< \boldsymbol{\lambda }\cdot \Delta \boldsymbol{u} \right> =\sum_{l=2}^{\infty}{\frac{\lambda _{\alpha _1}\lambda _{\alpha _2}\cdots \lambda _{\alpha _l}S_{\alpha _1\alpha _2\cdots \alpha _l}}{\left( l-1 \right) !}},
\end{align}
where we have used the Einstein summation convention for tensor indices, and
\begin{align}
S_{\alpha _1\alpha _2\cdots \alpha _l}=\partial _{(\alpha _1}\partial _{\alpha _2}\cdots \partial _{\alpha _{l-1}}u_{\alpha _l)}-\frac{1+\left( -1 \right) ^l}{2}\frac{\delta _{(\alpha _1\alpha _2}\delta _{\alpha _3\alpha _4}\cdots \delta _{\alpha _{l-1}\alpha _l)}}{l+1}\left( \nabla ^2\right) ^{\frac{l}{2}-1}\boldsymbol{\nabla }\cdot \boldsymbol{u}
\end{align}
may be recognized as a rank-$ l $ deformation tensor, which reduces to the ordinary traceless shear rate $ S_{\alpha\alpha '} $ for $ l=2 $. Note that the round brackets in indices denote symmetrization (the sum of all $ l! $ permutations divided by $ l! $), and
\begin{align}
\frac{\left< \lambda _{\alpha _1}\lambda _{\alpha _2}\cdots \lambda _{\alpha _{l}} \right>}{\lambda ^{l}}=\frac{1+\left( -1 \right) ^{l}}{2}\frac{\delta _{(\alpha _1\alpha _2}\delta _{\alpha _3\alpha _4}\cdots \delta _{\alpha _{l-1}\alpha _l)}}{l+1}.
\end{align}

On the other hand, the $ \ell $th term of the spherical harmonic expansion defines an irreducible tensor of rank $ \ell $, which is symmetric and vanishes when contracted on any pair of indices, with $ 2\ell +1 $ independent components \citep{1975ctf..book.....L}. This is however not the property of $ S_{\alpha _1\alpha _2\cdots \alpha _l} $ except for $ l=2 $, or the case of $ \boldsymbol{\nabla}\cdot\boldsymbol{u}=0 $ (incompressible flow) and $ \nabla ^2\boldsymbol{u}=\boldsymbol{0} $, implying that the Cartesian and spherical descriptions are generally inequivalent. Moreover, the case in which the two descriptions are equivalent is incompatible with our problem of interest, as turbulence propagation is inhibited by the Laplace equation \citep{1987flme.book.....L}.
\section{Other Models}\label{appxb}
In Fig.~\ref{f1}, the A14 model is given by Eq.~(21) in \citet{2014PhRvL.112b1101A}:
\begin{align}
C_{\ell}=\frac{18C_1}{\left( \ell +1 \right) \left( \ell +2 \right) \left( 2\ell +1 \right)}.
\end{align}
The AM15 model corresponds to Eq.~(16) in \citet{2015ApJ...815L...2A}:
\begin{align}
C_{\ell}=\frac{2\left[ \ell \left( \ell +1 \right) +q\left( q-1 \right) \right] \Gamma \left( 4-q \right) \Gamma \left( \ell +q-1 \right)}{\ell \left( \ell +1 \right) \left[ 2+q\left( q-1 \right) \right] \Gamma \left( \ell -q+3 \right) \Gamma \left( q \right)}C_1,
\end{align}
where $ q $ is defined via the relative scattering rate $ \nu _\text{r}\propto \left( 1-\cos\theta \right) ^q $, with $ \theta $ the angle between directions of motion of two particles. The fitting result is $ q\approx 0.6 $, implying $ C_\ell\propto\ell ^{-2.8} $ for $ \ell\gg 1 $ in AM15.

As $ C_1 $ is believed to be dominated by the diffuse dipole anisotropy, the energy dependence of $ C_\ell $ in A14 and AM15 can be affected largely by the distribution of CR sources, which is not the subject of this study.
\end{appendices}
\bibliographystyle{aasjournal}
\bibliography{ref}
\end{document}